\documentclass{PoS}

\usepackage{xspace}
\usepackage{amsmath}
\usepackage{wrapfig}

\def\CP      {\ensuremath{C\!P}\xspace}

\def\CPT     {\ensuremath{C\!PT}\xspace}

\def\P{\ensuremath{P}\xspace}
\def\T{\ensuremath{T}\xspace}

\def\Pz{\ensuremath{P^0}\xspace}
\def\Pzb{\ensuremath{\overline{P}^0}\xspace}

\def\Dz{\ensuremath{D^0}\xspace}
\def\Dzb{\ensuremath{\overline{D}^0}\xspace}

\def\Bz{\ensuremath{B^0}\xspace}
\def\Bzb{\ensuremath{\overline{B}^0}\xspace}

\usepackage{relsize}
\def\babar{\mbox{\slshape B\kern-0.1em{\smaller A}\kern-0.1em
    B\kern-0.1em{\smaller A\kern-0.2em R}}\xspace}

\title{\T symmetry invariance tests in neutral meson decays}

\ShortTitle{\T symmetry invariance tests in neutral meson decays}

\author{\speaker{Adrian BEVAN}%
        Queen Mary University of London\\
        E-mail: \email{a.j.bevan@qmul.ac.uk}}

\abstract{We outline how the time-reversal symmetry \T can be systematically used to test the Kobayashi-Maskawa mechanism
          embedded in the CKM matrix using pairs of $B$ mesons created at the $\Upsilon(4S)$ and pairs of $D$ mesons 
          from $\psi(3770)$. }

\FullConference{The European Physical Society Conference on High Energy Physics \\
   18-24 July, 2013\\
    Stockholm, Sweden}

\begin{document}

Weak decays are known to violate the set of improper (i.e. discrete) transformations $P$ (parity), $C$ (charge conjugation), 
$T$ (time-reversal), and $CP$.  
It is well known that electromagnetism and the strong force obey these symmetries, as bound by existing 
experimental data. The combination $CPT$ is related to the Lorentz group, and is conserved in locally gauge 
invariant quantum field theories.  Models of quantum gravity may violate Lorentz invariance, and hence \CPT.  These
violations in turn could manifest themselves as a difference between the \CP and \T violations found in weak decays,
hence \CP, \T, and \CPT tests are complementary parts of a triplet of tests of quark flavour transitions.
It has long been known that one can perform a Kabir asymmetry
test that can be interpreted in a dual way, i.e. as a \CP and \T symmetry test~\cite{Kabir:1970ts}. In 2012 the \babar experiment,
following a prescription laid down in~\cite{Banuls:1999aj}, performed a test of \T symmetry non-invariance (as well as \CP and \CPT tests) 
in $B$ decays using pairs of mesons, one decaying via a flavour filter, and the other decaying into either the \CP-even or \CP-odd 
filter state corresponding to a golden mode decay $B\to c\overline{c} K^0_{S,L}$~\cite{Lees:2012kn}.  These proceedings summarise
the work presented in~\cite{Bevan:2013rpr}, where we discuss applications of this methodology to other sets of
filter bases in the context of entangled pairs of neutral $B_{d,s}$ and $D$ mesons.  The remainder of these proceedings
start with a recap of the methodology, followed by considerations of applications to $B$ and $D$ decays before concluding
with a brief summary.

%
%
Entangled pairs of neutral pseudoscalar mesons (denoted by $P_1$ and $P_2$) provide one with access to a physical system were 
the time ordering of events can be naturally reversed by comparing the two time orderings of the 
superposition $\Phi$.  The wave function associated with such an entangled state 
is $\Phi = (|P_1 P_2\rangle + |P_2 P_1\rangle)/\sqrt{2}$, and this can  be experimentally tested as there is a 
pair of mesons that collapse into either the first or the second time ordering.  The next 
key point is to understand the \T conjugate pairs of decay filters required to experimentally
reconstruct and compare between the two time orderings.  In order to unambiguously test \T one requires
two different orthonormal basis pairs.  These can be any orthonormal pairs, but for convenience flavour 
i.e. $\{particle, \, antiparticle\}$ and \CP eigenvalue $\{+, \, -\}$ are obvious choices.  Hence 
there are four distinct comparisons that can be made for a given scenario. These are 
(i) $\Pzb\to \P_-$ vs $\P_- \to \Pzb$,
(ii) $\P_+\to \Pz$ vs $\Pz \to \P_+$,
(iii) $\Pzb\to \P_+$ vs $\P_+\to \Pzb$,
(iv) $\P_-\to \Pz$ vs $\Pz\to \P_-$, 
where \Pz (\Pzb) refers to a neutral meson flavour filter decay identifying a particle or anti-particle, and those
with subscripts $\pm 1$ refer to the eigen value of the \CP filter final state.

The flavour filter basis pair is defined by neutral meson decays to flavour specific final states, i.e. states
that are only accessible to particle or anti-particle decays.   The definition of this set could be extended to 
include Cabibbo allowed vs Cabibbo suppressed decays in the case of flavour tagging for $D$ mesons, at the
cost of dilution of the signal. The original proposal for the \CP filter basis was to use the approximately
orthonormal set given by the $B$ decay to a charmonium ($c\overline{c}$) plus a $K_S$ ($\CP = -1$) or $K_L$ ($\CP = +1$).  This
can be extended to include a number of other final states as discussed below.  As pointed out in~\cite{Bevan:2013rpr}, 
one should also recognise that decays of pseudoscalars to two spin one (vector or axial-vector) particles also constitute a set
of exact \CP filter basis pairs if one performs a full angular analysis to separate out the even and odd components.
This broadens the range of \T violation tests that one can perform in the Standard Model.

%
%
Interactions resulting in the decay of $\Upsilon(4S) \to \Bz\Bzb$ and $\psi(3770)\to \Dz\Dzb$ are equivalent in terms
of quantum entanglement of the final state.  This entanglement has been validated by Belle for the $\Upsilon(4S)$ 
scenario~\cite{Go:2007ww} and is assumed to be valid for $\psi(3770)$ decays (which can be tested at a suitable 
charm factory).  

As detailed in~\cite{Bevan:2013rpr} it is possible to measure the Unitarity Triangle angle $\beta$ ($\alpha$)
using $b\to c$, $s$, $d$ ($u$) decays.  In addition one can perform \T symmetry tests in $c\to u$, $d$, $s$ 
transitions at the $\psi(3770)$.  For charm the goal is first to test \T violation in mixing, and ultimately 
one day to extend the interpretation to constraining the charm Unitarity Triangle angle $\beta_c$.
Interpretation is in terms of the \T violating phase measured for $\lambda_f = (q/p)(\overline{A}/A)$.  The step
from mixing constraints $(q/p)$ to the weak structure of interference between mixing and decay $(\lambda_f)$ requires 
precision that goes beyond the current generation of experiments, and an improved understanding of hadronic 
uncertainties in the charm sector.  Constraints on $\gamma$ using $c\to u$ decays are not viable as, while these 
processes contribute to a number of decays, they appear as sub-dominant penguin amplitudes.  A review of
possible interesting modes to construct \CP filter bases from can be found in~\cite{Bevan:2013rpr,Bevan:2011up}.

\begin{wraptable}{r}{0.45\textwidth}
\label{tbl:estimates}
\caption{Estimated sensitivities on $\sigma(2\sin 2\beta)$ for promising \CP filter pairs.}
\begin{tabular}{c|cc}
Filter basis pair & $B$ Factories & Belle II \\ \hline\hline
$\eta^\prime K^0_{S/L}$ & 0.6 & 0.08 \\
$\phi K^*$              & 1.1 & 0.13 \\
$\eta K^0_{S/L}$        & 1.8 & 0.17 \\
$\omega K^0_{S/L}$      & 2.0 & 0.22 \\
$D^* D^*$               & 2.0 & 0.29 \\ \hline
\end{tabular}
\label{exposure}
\end{wraptable}

A number of previously unthought of \CP filter basis pairs can be used to search for \T violation.  These 
include the following $B$ decays: the $b\to s$ loop decays $B \to (\eta^\prime, \phi, \omega) K_{S,L}$
and $B\to \phi K^*$; $b\to d$ loop transitions $B\to D^{*+}D^{*-}$; 
the $b\to c$ transition $B\to J/\psi K^*$; and $B\to J/\psi \rho$ which is a colour suppressed $b\to c$ transition with a
potential $b\to d$ penguin contamination, all these states measure $\beta$.
Additionally one can measure $\alpha$ using $b\to u$ transitions such as $B \to \rho\rho$ and $a_1\rho$ decays.  
Table 1 summarises estimates of precisions attainable for \T symmetry parameters (related
to $\pm 2\sin2\beta$) for $B$ decays at the current $B$ Factories and at Belle II.  Effects at the level of the Standard Model
expectation should be observable in all modes at Belle II (with 50$\mathrm{ab}^{-1}$ of data).
In terms of
$D$ decays the \CP filter basis modes of interest include: $\Dz \to  K^0_{S,L} (\omega, \eta, \eta^\prime, 
\rho^0, \phi, f^0, a_0)$ and kinematically allowed modes with pairs of (axial-)vector mesons in the final state.

%
%

In summary we outline a set of \T symmetry invariance tests of $b\to u$, $c$, $d$, and $s$ 
as well as $c\to d$, and $s$ filter basis transitions that would enable one to over-constrain
our understanding of the unitarity of the CKM matrix in terms of $B_{d,s}$ and $D$ decays in the
context of the Standard Model.  After almost five decades of \CP violation measurements it is possible to embark
upon an equivalent era of \T (and \CPT) violation tests in weak interactions to probe possible new physics contributions
in tree and loop decays.

\end{document}